\documentclass[a4paper,11pt]{article}%
\usepackage{graphicx}
\graphicspath{{Figures/}} 
\usepackage{amsmath}
\usepackage{amssymb}
\usepackage{subcaption}
\usepackage[colorlinks]{hyperref}
\usepackage{color}%
\usepackage{cite}
\usepackage{float}

\newcommand{\bra}{\begin{array}}
\newcommand{\era}{\end{array}}
\newcommand{\beq}{\begin{equation}}
\newcommand{\eeq}{\end{equation}}
\newcommand{\bqr}{\begin{eqnarray}}
\newcommand{\eqr}{\end{eqnarray}}

\def\BC{\bb C}
\def\_\BC{\bbi C}

\def\( {\left(}
\def\) {\right)}
\def\no2 {{\textstyle{n\over 2}}}

\begin{document}
\begin{titlepage}
\setcounter{page}{1}
\renewcommand{\thefootnote}{\fnsymbol{footnote}}

\begin{center}
{\Large \bf { 
Confining type-II spherical core-shell quantum dot heterostructures with narrow and wide band gaps
}}

\vspace{5mm}

{\bf T. Shelawati\footnote{\sf shelawati.tiansin@upm.edu.my}}$^{a}$,
{\bf M.S. Nurisya\footnote{\sf risya@upm.edu.my}}$^{a,b}$
and {\bf A. Jellal\footnote{\sf a.jellal@ucd.ac.ma}}$^{c,d}$

\vspace{5mm}
{$^{a}$\em Department of Physics, Faculty of Science, Universiti Putra Malaysia, }\\
{\em 43400 UPM Serdang, Selangor, Malaysia}

{$^{b}$\em Laboratory of Computational Sciences \& Mathematical Physics, \\  Institute for Mathematical Research, Universiti Putra Malaysia, 43400 UPM Serdang, Selangor, Malaysian}

{$^{c}$\em Laboratory of Theoretical Physics,  
				Faculty of Sciences, Choua\"ib Doukkali University},\\
			{\em PO Box 20, 24000 El Jadida, Morocco}

			{$^{d}$\em Canadian Quantum  Research Center,
				204-3002 32 Ave Vernon, \\ BC V1T 2L7,  Canada}

\vspace{30mm}
\begin{abstract}
Using a single-band model, the lowest transition energy was analysed between the lowest unoccupied molecular orbital (LUMO) of the conduction band and the highest occupied molecular orbital (HOMO) of the valence band. We focus on categorising the  confinement strength in type-II core-shell quantum dots (CSQDs) based on the step-potential and show how it will affect their transition energy. Our model is applied to 
CSQDs of the heterostructures PbS/CdS and ZnTe/ZnSe through 
narrow and wide band gaps, respectively.
It found that PbS/CdS CSQDs demonstrates a strong confinement in which their transition energy would increase more compared to its weak confinement case in ZnTe/ZnSe CSQDs. The weak confinement case also demonstrated both blue-shift and red-shift of photoluminescence emission compared to the bulk ZnTe and ZnSe for which it can be inferred as pseudo type-II CSQDs. This would help experimentalists to tune the transition energy of type-II model in order to fabricate photons with longer carrier lifetime compared to the type-I model.
\vspace{3cm}

\end{abstract}
\end{center}
\end{titlepage}

\section{Introduction}
\label{intro}
Quantum dots realize the limiting case
of size quantization in semiconductors, which cause  modifications
of their electronic properties  \cite{1}.
The colloidal core-shell quantum dots (CSQDs) are heterostructures and  can be
obtained by chemical synthesis
\cite{111, 222, 333} with reproducible and controllable size and shape and
low fabrication cost.
The demand and applications of nanotechnology in devices for biological tagging  \cite{jia2013, poulsen2017, ca2019, kim2003}, diodes \cite{ca2019}, solar cells \cite{jia2013, ca2019, vantakhah2013}, 
and lasers \cite{ca2019, kim2003, vantakhah2013} have increased over the years made quantum dots to be studied extensively, and developed further into core-shell structures. Through this chemical surface passivation, core-shell quantum dots (CSQDs) would have better photo-stability \cite{kostic2012} and a  highly tunable optical spectra \cite{kostic2012, cheche2013}. Capping a quantum dots surface with another semiconductor material would also give room for wavefunction engineering. Then it is possible to form CSQDs with type-I band alignments or type-II band alignments for the same pair of materials by manipulating the materials composition or by altering their environment by introducing electric field \cite{poulsen2017, nandan2019}.

While a type-I CSQDs would have both of their charge carriers confined within the same region, in type-II CSQDs, both electrons and holes would be confined in different regions; one within the core region and the other within the barrier interface region \cite{bimberg1999}. This novel ability to control the region of carriers confinement leads to lower non-radiative recombination rate \cite{dorfs2008} in type-II CSQDs. Type-II also offers longer carrier lifetimes while reducing the radiative recombination probability \cite{rocha2019, shoji2017}. In addition, type-II CSQDs also preferred in engineering a near-infrared or infrared photons that can be used to assist surgery for direct visual guidance by injecting the photons into the skin \cite{kim2003b}.

Due to these promising features of type-II CSQDs, it is best to acknowledge the importance to understand the effects of step-potentials in type-II CSQDs by using the single band model and see how it will affect their transition energy. This is much due to the previous findings on how different confinement strength in type-I CSQDs affects their transition energy \cite{shela2019}. Previous works have tackled the transition energy problem in type-II CSQDs using continuum mechanical model (CMM), valence force field model (VFF) and effective mass approximation (EMA) \cite{stier1999, kaledin2018, tyrrell2011}. However, CMM and VFF approaches involved 8-band \textbf{k$\cdot$p} method while EMA often involved with (2,8)-band Hamiltonian approach which often hard to solve and require longer calculations compared to the single-band model \cite{stier1999, tyrrell2011}. As the name suggests, the single-band model only focusing on the transition energy between the lowest unoccupied molecular orbital (LUMO) of conduction band and the highest occupied molecular orbital (HOMO) of  valence band.

The present  work reports the computation of the lowest transition energy between the lowest unoccupied molecular orbital of the conduction band and the highest occupied molecular orbital of the valence band. This will be done by  using a single band model involving the confining potentials for electrons and holes. The solutions of the energy spectrum  will be determined by requiring
 separable wave functions in radial and angular parts. The radial part will involve the physical parameters generated from the confinements. 

The paper is organized as follows.
Section \ref{S2} would be dedicated for discussion on the theoretical single-band approach on energy transition observed in CSQDs. In section \ref{redis}, the mathematical formulation of the single-band theory is applied on type-II CSQDs for strong and weak confinement cases for which analogous to the CSQD heterostructure with narrow and wide band gap. The analysis of their radial probability will be presented in section \ref{S3}. We close by giving   conclusion remarks.

\section{Theory\label{S2}}

\begin{figure}[!ht]
  \centering
  \includegraphics[scale=0.5]{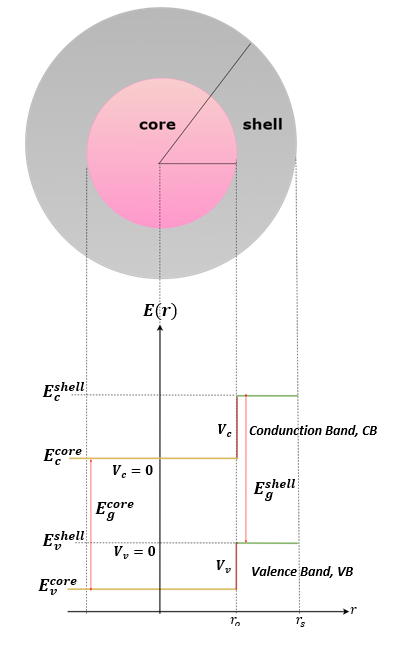}
  \caption{(color online) Schematic diagram of band line-up of Type-II CSQDs.}
  \label{fig:bandst2}
\end{figure}

We consider type-II
core-shell quantum dots (CSQDs) based on the step-potential. 
Figure \ref{fig:bandst2} shows a schematic diagram for type-II CSQDs, with $E_{c(v)}$ and  $V_{c(v)}$ are the energy and step-potential for conduction(valence) band, while $r_o$ and $r_s$ are the radius of core quantum dots and CSQDs respectively. For carriers outside the shell layer, the corresponding $V(r)$ would be infinite. This assumption of having an infinite barrier would lead to a complete spatial separation between the CSQDs and its environment. Therefore the wavefunction overlap and the Coulomb potential outside the ``artificial atom'' are negligible  \cite{bimberg1999}. 

To achieve our goal, let us 
introduce the step-potential due to electrons confinement \cite{shela2019}
\begin{align}\label{potc}
V_e(r)=\left\{\begin{matrix}
0   &; &0\leq r\leq r_o\\ 
V_{c} &; &r_o\leq r\leq r_s\\
\infty &; &r\geq r_s .
\end{matrix}\right.
\end{align}
and focus only 
interesting solution comes from solving the radial part $R_{nl}(r)$ of the following three-dimensional wavefunction 
\begin{equation}
\label{eqn:separate}
\Psi_{nlm}(\vec{r})=R_{nl}(r)Y_{lm}(\theta,\phi),
\end{equation}
where  $Y_{lm}$ are its spherical harmonic functions, 
with $n$, $l$ and $m$ are the principal, orbital and magnetic quantum numbers respectively. Since this work is focusing on analysing a single-band model, hence one only needs to consider the transition energy between the lowest unoccupied molecular orbital (LUMO) of the conduction band and the highest occupied molecular orbital (HOMO) of valence band, that is taking $n=1$ and $l=m=0$. According the potential profile Eq. \eqref{potc}, 
the radial wavefunction for ground state electron then would be
\begin{align}
\label{eqn:strongcon2e}
R_i(r)=\left\{\begin{matrix}
R_1(r) =A J_l(k_1r)   &; & 0\leq r\leq r_o\\ 
R_2(r) =B H_l(k_2r)+C N_l(k_2r) &; & r_o\leq r\leq r_s\\
R_3(r) =0 &; & r\geq r_s.
\end{matrix}\right.
\end{align}
where $j_l(k_i)$, $h_l(k_i)$ and $n_l(k_i)$ are spherical Bessel function of the first kind, modified  spherical Bessel function of the first kind and modified spherical Bessel function of the second kind respectively, with $i=1$ for core, $i=2$ for shell and $i=3$ for outside shell layer. The involved wave vector for both regions are
\begin{equation}
 k_1=\frac{\sqrt{2m^e_1E_e}}{\hbar},\qquad  k_2=\frac{\sqrt{2m^e_2[V_c(r)-E_e]}}{\hbar}
 \end{equation}
  and  $m_i^e$ is the effective mass of electrons. 
Eq. \eqref{eqn:strongcon2e} would form a transcendental equation that obey  det$(E_{e})=0$ and thus
\begin{equation}
\label{eqn:trans2e}
\dfrac{m_2^*}{m_1^*}\dfrac{J'_0(k_1r_o)}{J_0(k_1r_o)}=\dfrac{H'_0(k_2r_o)n_0(k_2r_s)- H_0(k_2r_s) N'_0(k_2r_o)}{H_0(k_2r_o)n_0(k_2r_s)- H_0(k_2r_s) N_0(k_2r_o)}.
\end{equation}
One will obtain the energy of electrons, $E_{e}$ after solving Eq. \eqref{eqn:trans2e}. As for these electron states, the numerical analysis is the same as solving for electron states in Type-I CSQDs, as their electrons are confined within the conduction band of the core region.

In the case of Type-II, the confinement for holes would be no longer within the valence band of core region, but instead, it would be spatially confined within the valence band of both core and shell regions. This would reflect different step-potential distribution compared to their electron counterpart, where they only confined within their core region as
\begin{align*}
V_h(r)=\left\{\begin{matrix}
V_{v}   &; &0\leq r\leq r_o\\ 
0 &; &r_o\leq r\leq r_s\\
\infty &; &r\geq r_s 
\end{matrix}\right.
\end{align*}
and then the corresponding radial wavefunctions for ground state of holes are
\begin{align}
\label{eqn:strongcon2h}
R_i(r)=\left\{\begin{matrix}
R_1(r) =D T_l(k'_1r)   &; & 0\leq r\leq r_o\\ 
R_2(r) =F P_l(k'_2r)+G Q_l(k'_2r)) &; & r_o\leq r\leq r_s\\
R_3(r) =0 &; & r\geq r_s,
\end{matrix}\right.
\end{align}
where $t_l(k_i)$, $p_l(k_i)$ and $q_l(k_i)$ are modified spherical Bessel function of the first kind, spherical Bessel function of the first kind and spherical Bessel function of the second kind, respectively. The wave vectors are
\begin{equation}
 k'_1=\frac{\sqrt{2m^h_1[V_c(r)-E_h]}}{\hbar},\qquad  k'_2=\frac{\sqrt{2m^h_2E_h}}{\hbar}
 \end{equation}
  and  $m_i^h$ is the effective mass of holes.
Also here 
Eq. \eqref{eqn:strongcon2h} will form the transcendental equations that can be written as 
\begin{equation}
\label{eqn:trans2h}
\dfrac{m_2^h}{m_1^h}\dfrac{T_0(k'_1r_o)}{T_0(k'_1r_o)}=\dfrac{P'_0(k'_2r_o)Q_0(k'_2r_s)- P_0(k'_2r_s) Q'_0(k'_2r_o)}{P_0(k'_2r_o)Q_0(k'_2r_s)- P_0(k'_2r_s) Q_0(k'_2r_o)},
\end{equation}
which would give the numerical value for energy of the holes, $E_h$. 

Combining all solutions, the lowest transition energy, $E_\text{qd}$ for CSQDs can be solved by using the Brus equation~\cite{brus1986}
\begin{equation}
\label{eqn:total}
E_\text{qd}=E_g^{bulk} + E_e + E_h -\frac{1.8 e^2}{4 \pi \varepsilon_o \varepsilon_r r_o},
\end{equation}
where $e$ is the elementary charge, $\epsilon_o$ is the permittivity of free space, $\epsilon_r$ is the relative permittivity and $r_o$ is the core radius.

\section{Results and discussions}\label{redis}
\subsection{Type-II CSQDs with strong confinement}

Now the obtained mathematical formulation will be applied to a type-II CSQDs with narrow core band gap that encapsulated with a shell layer with much wider band gap. To imply a strong confinement case, the band alignment is chosen such that $V_{c} \gg E_g^{bulk}$, and $V_{v} \approx \frac {1}{2}E_g^{bulk}$. For this purpose, we choose PbS/CdS core-shell QDs with material parameters as shown in Table \ref{table:2}. 

\begin{table}[!htbp] 
\centering
 \caption{\sf List of parameters for PbS/CdS core-shell QDs.}
 \begin{tabular} {|c |c |c |} 
 \hline 
  Materials & PbS \cite{mishra2008} & CdS  \cite{mishra2008, madelung1999} \\
 \hline
 Band gap (eV) & 0.41 &   2.5 \\ [5pt]\hline
 $m_e/m_0$ & 0.08 &  0.2 \\ [5pt] \hline
$m_h/m_0$ & 0.08 &  0.7  \\ [5pt] \hline
Relative permittivity, $\varepsilon_r$ & 169 &  8.73  \\ [5pt] \hline
$V_c$ (eV) & \multicolumn{2}{|c|}{2.29}  \\ [5pt] \hline
$V_v$ (eV) & \multicolumn{2}{|c|}{0.20}   \\ [5pt] \hline 
 \end{tabular}
\label{table:2}
\end{table}

\begin{figure}[h!] 
\centering
\includegraphics[scale=0.4]{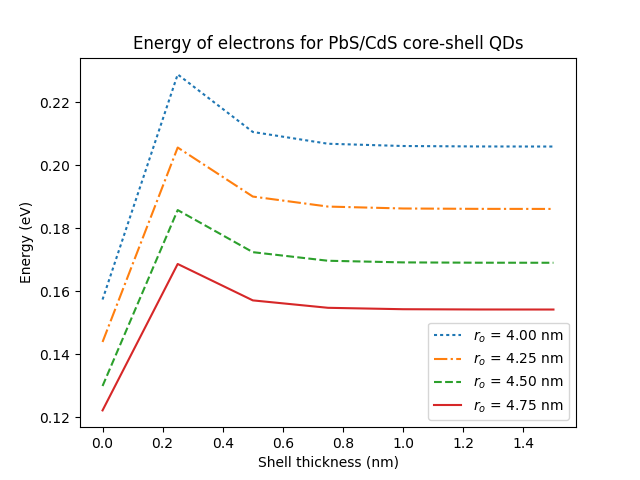}\ \ \ \ \ 
 \includegraphics[scale=0.4]{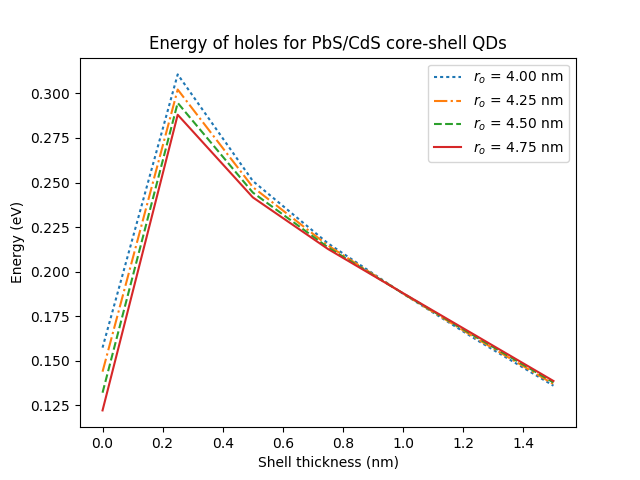}
\includegraphics[scale=0.4]{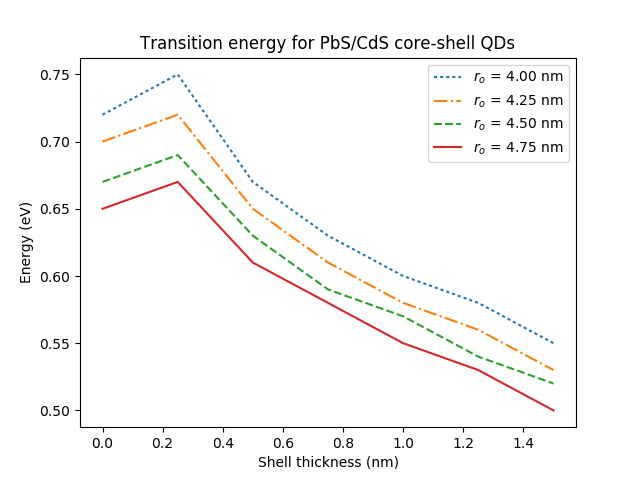}
\caption{\sf (color online) Energy of electrons (left panel) and  holes (right panel) for  PbS/CdS. Transition energy for PbS/CdS CSQDs (bottom panel).}
\label{fig:pbscds3}
\end{figure}

Using Eqs. \eqref{eqn:trans2e}, \eqref{eqn:trans2h} and \eqref{eqn:total}, the energy of electrons ($E_e$), holes ($E_h$) and the total transition energy in the type-II CSQDs ($E_\text{qd}$), we present their behaviors  
in Figure \ref{fig:pbscds3}.

%
%

\newpage
\subsection{Type-II CSQDs with weak confinement}

We apply  our model single-band model  to a wide core QDs, encapsulated with a shell layer with slightly wider band gap. Then, to insinuate a weak confinement case, the step-potential is chosen such that $V_{c} \gtrapprox \frac {1}{2}E_g^{bulk}$, while keeping $V_{v} \approx \frac {1}{2}E_g^{bulk}$. Here we choose ZnTe/ZnSe core-shell QDs with material parameters as shown in Table \ref{table:3}.\\

\begin{table}[!htbp] 
\centering
 \caption{\sf List of parameters for ZnTe/ZnSe core-shell QDs.}
 \begin{tabular} {|c |c |c |} 
 \hline
 Materials & ZnTe  \cite{jia2013, mishra2008} & ZnSe \cite{jia2013,mishra2008} \\
 \hline
 Band gap (eV) & 2.394 &   2.8215 \\ [5pt]\hline
 $m_e/m_0$ & 0.11 &  0.14 \\ [5pt] \hline
$m_h/m_0$ & 0.7 &  0.6  \\ [5pt] \hline
Relative permittivity, $\varepsilon_r$& 8.7 &  9.1  \\ [5pt] \hline
$V_c$ (eV) & \multicolumn{2}{|c|}{1.6275}  \\ [5pt] \hline
$V_v$ (eV) & \multicolumn{2}{|c|}{1.2000}   \\ [5pt] \hline
 \end{tabular}
\label{table:3}
\end{table}

\begin{figure}[h!] 
\centering
\includegraphics[scale=0.4]{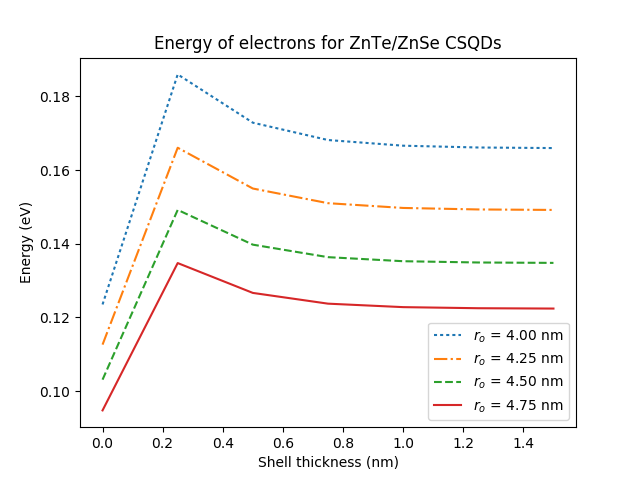}\ \ \ \ \
\includegraphics[scale=0.4]{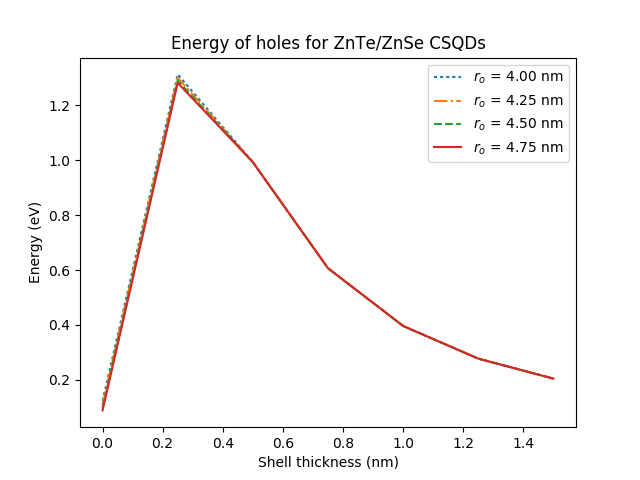}\label{fig:znteznse2}
\includegraphics[scale=0.4]{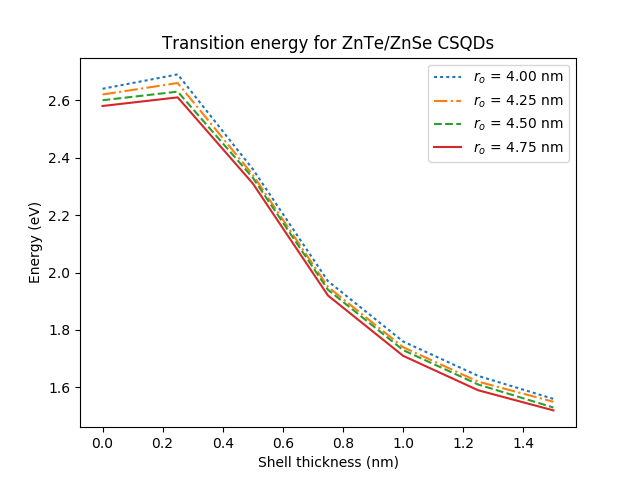}\label{fig:znteznse3}
\caption{\sf (color online) Energy of electrons (left panel) and  holes (right panel) for ZnTe/ZnSe. Transition energy for ZnTe/ZnSe CSQDs (bottom panel).} 
\label{fig:znteznse1}
\end{figure}

As before from  Eqs \eqref{eqn:trans2e}, \eqref{eqn:trans2h} and \eqref{eqn:total}, the energy of electrons ($E_e$), holes ($E_h$) and the total transition energy in the type-II CSQDs ($E_\text{qd}$)  for weak confinement case are presented in Figure \ref{fig:znteznse1}.

%
%

\section{Radial probability\label{S3}}

\subsection{Strong confinement case}

\begin{figure}[h!] 
\centering
\includegraphics[scale=0.5]{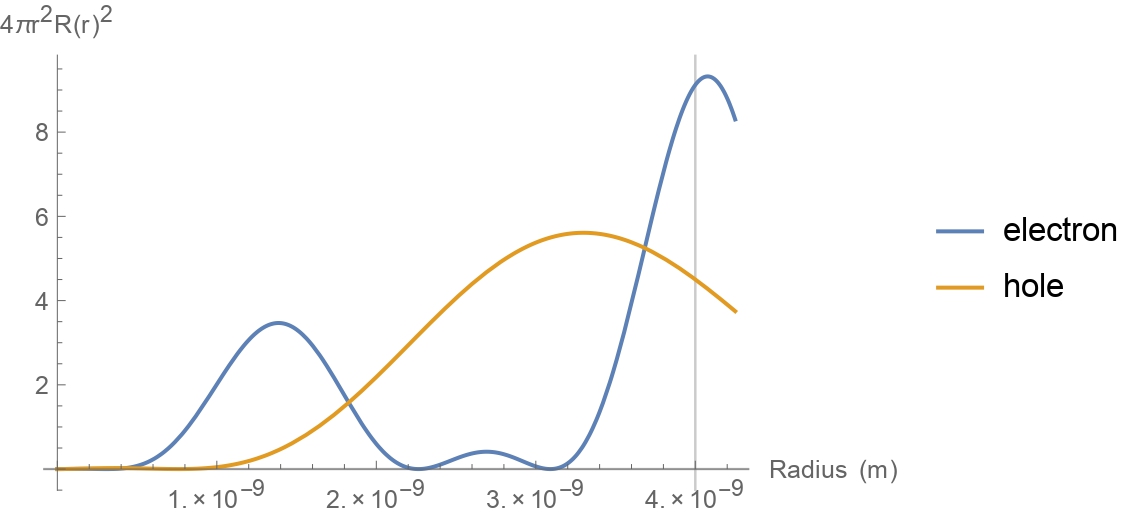}
\caption{\sf(color online) Radial probability of carriers in 4.25 nm PbS/CdS with 0.25 nm shell thickness.}
\label{fig:csPbS1}
\end{figure}

\begin{figure}[h!] 
\centering
\includegraphics[scale=0.5]{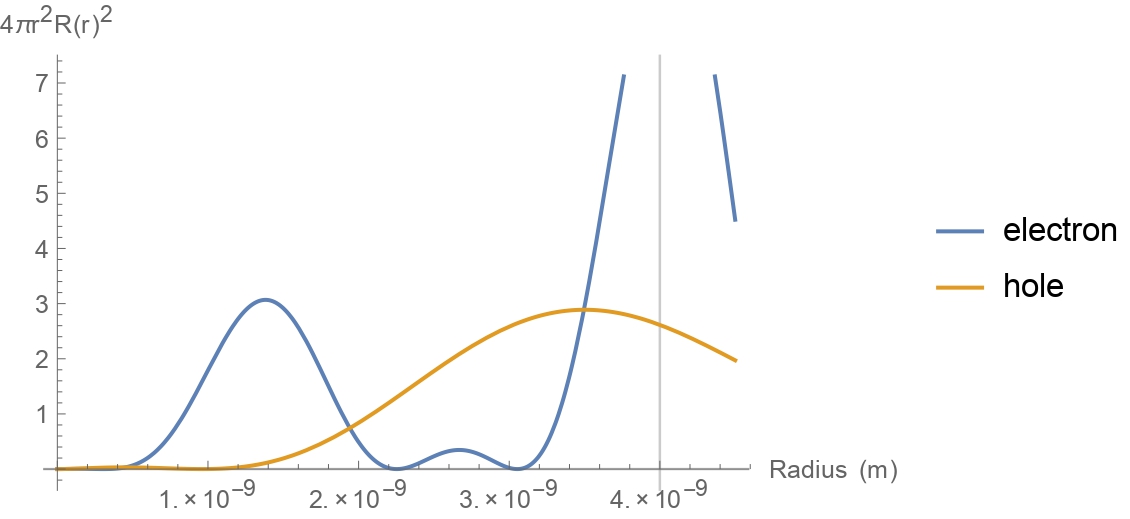}
\caption{\sf (color online) Radial probability of carriers in 4.50 nm PbS/CdS with 0.5 nm shell thickness.}
\label{fig:csPbS2}
\end{figure}

The overall trend for the transition energy in PbS/CdS CSQDs shows a jump around 5$\%$ compared to bare PbS quantum dots when we added 0.25 nm of shell thickness. However, it started to be decreasing after thicker shell layer applied, as shown in bottom panel of Figure \ref{fig:pbscds3}. This proves that capping a quantum dots into a core-shell form will cause stronger localization while inducing stress onto the charge carriers that will cause further lattice mismatch \cite{gong2014}. The induced lattice mismatch now responsible to minimize the strain caused by the strongly confined carriers. 
However, the trend of energy of electrons keeps constant although the shell thickness was increased for any given core radii. This is most probably due to the fact that the electrons is now in the state of highly localized, which makes them to be less ``triggered" by the change in shell thickness. This can be proven by comparing the radial probability of electrons in Figures \ref{fig:csPbS1} and \ref{fig:csPbS2} where different shell thickness did not affect on how the electrons are localized.

As for the holes, a jump in energy was shown in right panel of Figure \ref{fig:pbscds3},
before they started decreasing after 2.5 nm of CdS shell. As the strength of confinement is much less in the valence band compared to in conduction band, the holes are less likely to be highly localized as their electrons counterpart, making them to be ``aware'' of any change in shell thickness. Hence, as thickness is increased, the holes started to tunnel through between core and shell regions, as shown by their spread of radial probability in Figure \ref{fig:csPbS2}. It can be seen that the spread of radial wavefunction of holes are higher in Figure \ref{fig:csPbS2} compared to holes in \ref{fig:csPbS1}.

\subsection{Weak confinement case}

The total transition energy in ZnTe/ZnSe CSQDs is higher compared to their bare ZnTe quantum dots,  by almost 2$\%$, as shown in bottom panel of  Figure \ref{fig:znteznse1}. This lower jump is due to a weaker confinement strength compared to the case of PbS/CdS in previous section. The trend of energy of electrons in ZnTe/ZnSe is the same as in PbS/CdS, where their localized state causing them to be less influenced by the increasing of shell thickness. 
Unlike in the case of strong confinement where different core radii had different energy of holes, in weak confinement, the difference in energy of holes are hardly noticeable. The reason for this can be studied by observing their radial probability. 

\begin{figure}[h!] 
\centering
\includegraphics[scale=0.5]{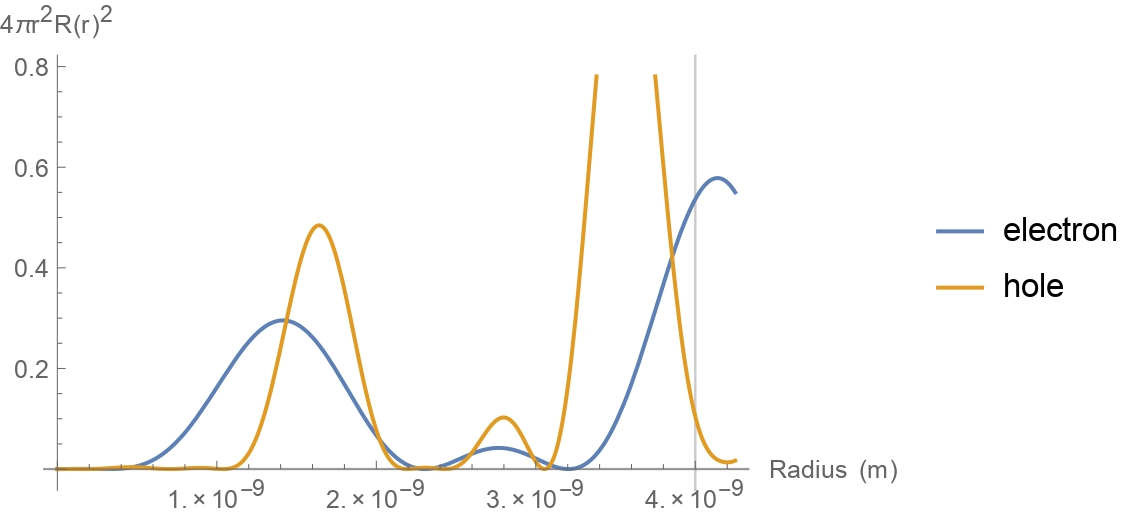}
\caption{\sf (color online) Radial probability of carriers in 4.25 nm ZnTe/ZnSe with 0.25 nm shell thickness.}
\label{fig:csznte1}
\end{figure}

\begin{figure}[h!] 
\centering
\includegraphics[scale=0.5]{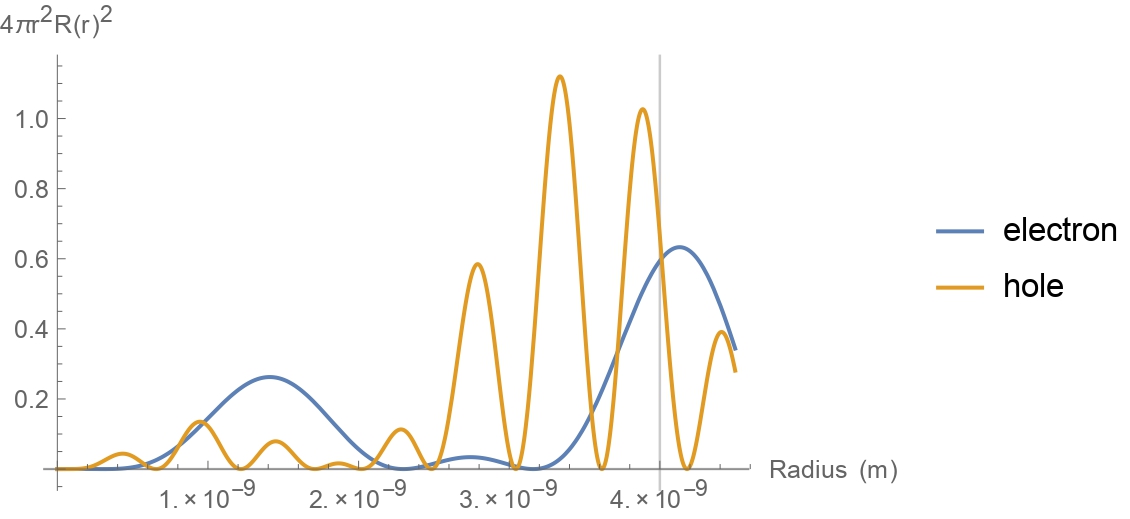}
\caption{\sf (color online) Radial probability of carriers in 4.50 nm ZnTe/ZnSe with 0.5 nm shell thickness.}
\label{fig:csznte2}
\end{figure}

Although the ZnTe/ZnSe CSQDs has type-II band alignments, it mimics the behaviour of type-I CSQDs where both charge carriers are likely to be confined within the same region. It can be seen in Figures \ref{fig:csznte1} and  \ref{fig:csznte2} that their holes are likely (not entirely) to be localized within the core region instead of within the shell band like any typical type-II CSQDs.  From these results, it can be inferred that weak confinement of the type-II case can be referred as pseudo-type-II CSQDs \cite{nandan2019}.

\subsection{Effect of core size}

\begin{figure}[h!] 
\centering
\includegraphics[scale=0.47]{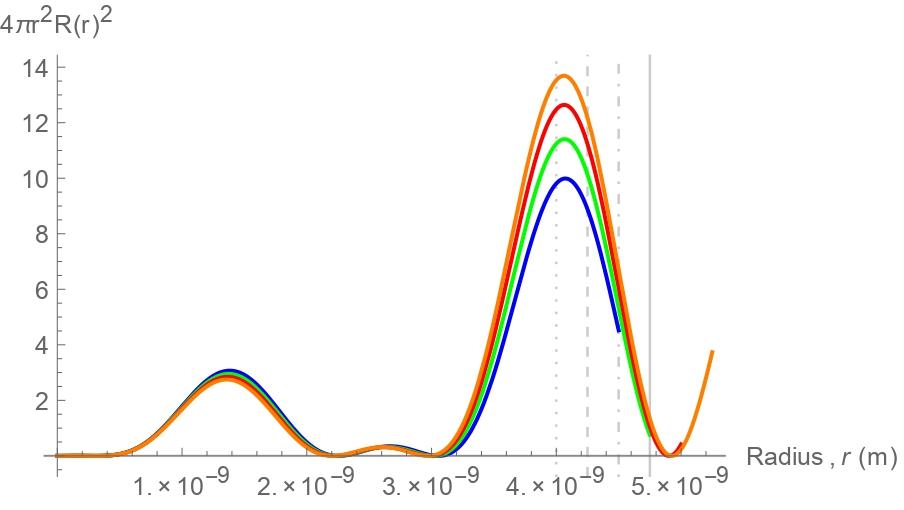}\ \ \
\includegraphics[scale=0.47]{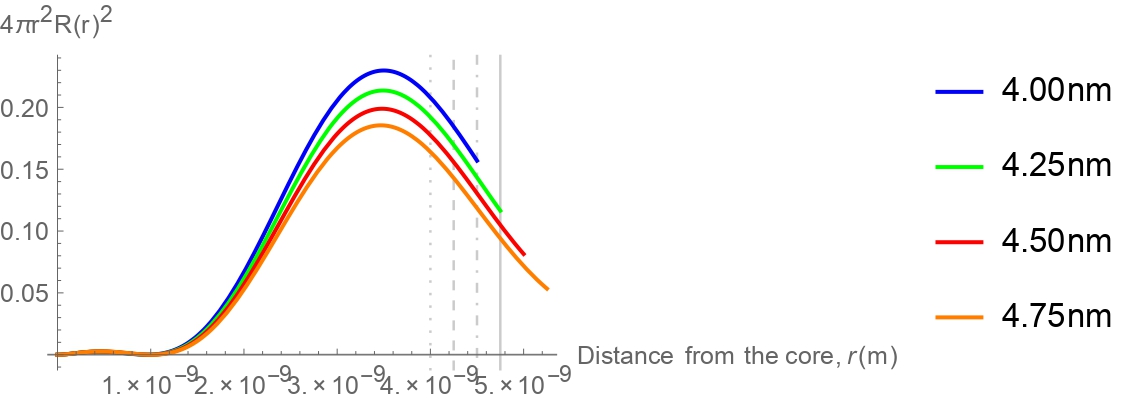}
\caption{\sf (color onine) Radial probability of electrons (left panel) and holes (right panel) in PbS/CdS CSQDs with 0.5 nm of shell thickness (strong confinement case).}
\label{fig:pic1}
\end{figure}

The effects of different core sizes is further studied by analysing their radial probabilities with the same shell thickness. 
In the case of strong confinement of PbS/CdS CSQDs, we can see from bottom panel of Figure \ref{fig:pbscds3} that the smallest CSQDs has the highest energy compared to the bigger ones. As expected, this small size of CSQDs contributes to a greater localization of charge carriers as shown in Figure \ref{fig:pic1}, where carriers inside 4.00 nm core size has the highest probability of being at the centre of the core. 

\begin{figure}[h!] 
\centering
\includegraphics[scale=0.47]{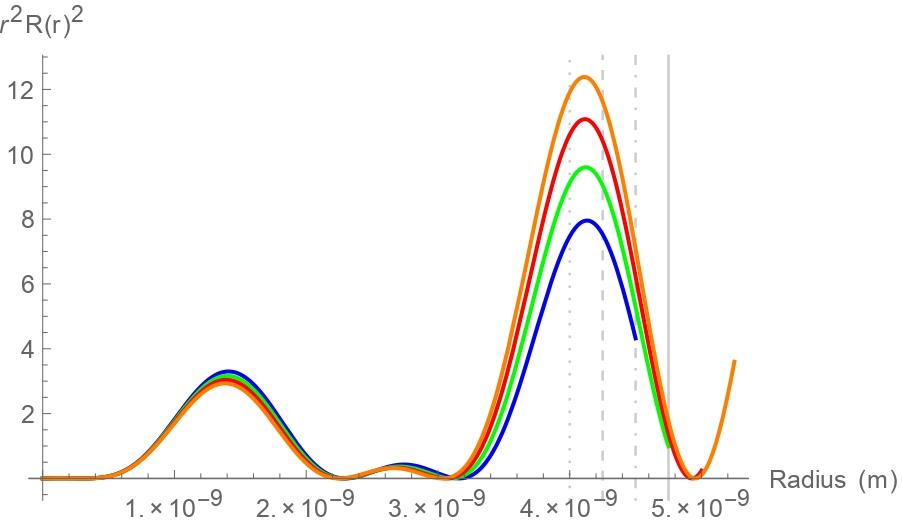}\ \ \
\includegraphics[scale=0.47]{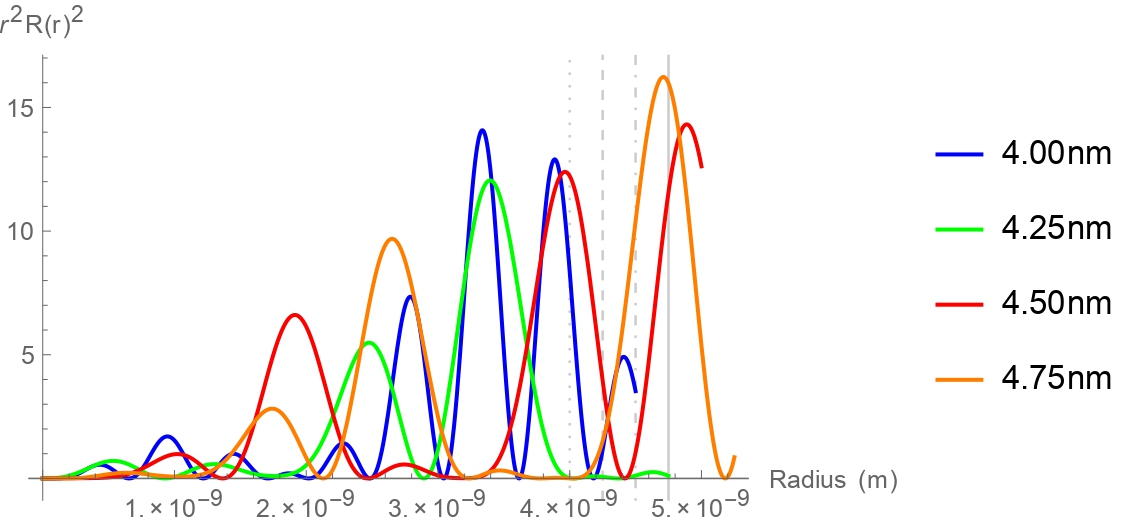}
\caption{\sf (color onine) Radial probability of electrons (left panel) and holes (right panel) in ZnTe/ZnSe CSQDs with 0.5 nm of shell thickness weak confinement case).} 
\label{fig:pic2}
\end{figure}

However, in the case of weak confinement of ZnTe/ZnSe CSQDs, the shell thickness only influence the electrons confinement, and not the holes. We also found that with different core radii, the holes still mostly spread throughout the core region as observed in Figure  \ref{fig:pic2}. Hence, both Figure \ref{fig:csznte2} and \ref{fig:pic2} confirm that core and shell sizes will not influence on how the holes will be localized or confined.

\section{Conclusion}
We have properly defined the strength of confinement  based on their step-potential rather than their size. The new conditions of strength confinement was justified by investigating the effects of strong and weak confinement cases in the type-II CSQDs with varying both core radius and shell thickness. It can be seen that by introducing strong confinement in CQDS, the transition energy would increase as high as 5$\%$, compared to the weak confinement case, where the increase of their transition energy is only around 2$\%$. From the theoretical point of view, this results demonstrate the flexibility to manipulate the combination of core and shell materials while addressing the strong and weak confinement conditions. Furthermore, the weakly confined charge carriers in type-II CSQDs has the ability to demonstrate red-shift emission where the transition energy of this CSQDs is lower compared to both of ZnTe and ZnSe band gaps. This  is favourable in fabricating near-infrared or infrared photons emission with longer carriers lifetime \cite{kim2003}. 

 Another worthy finding is how weak confinement case in type-II mimics confinement behaviour of type-I CSQDs where both charge carriers are confined within the same region, with their holes being less influenced by the changes in both core and shell sizes. This has brought us to refer such case as pseudo-type-II CSQDs. It is hoped that our findings can help experimentalist to further understand the reason behind behavioural changes in type-II core-shell quantum dots, which will help them to explain and justify their experimental findings.

\section*{Acknowledgments}

We would like to thank Prof Zainal Abidin Talib for his kind guidance in preparing this manuscript. This research was supported by Ministry of Higher Education Malaysia (MOHE) through the Fundamental Research Grant Scheme (FRGS/1/2018/STG02/UPM/02/10).

\end{document}